\documentclass[prb,twocolumn,showpacs]{revtex4}
\usepackage{amsmath}
\usepackage{graphics}
\usepackage{epsfig}
\usepackage{color}
\usepackage{amsfonts,amsmath,mathrsfs}
\usepackage{slashed}
\usepackage{hyperref}
\usepackage{epsfig}
\usepackage{latexsym}
\usepackage{bbm}
\usepackage{amssymb}
\usepackage{graphicx}
\usepackage{txfonts}
\usepackage{dcolumn}
\usepackage{bm}

\begin{document}
\title{Sign Structure, Electron Fractionalization, and Emergent Gauge Description of the Hubbard Model}
\author{Long Zhang}
\affiliation{Institute for Advanced Study and Collaborative Innovation Center of Quantum Matter, Tsinghua University, Beijing, 100084, China}
\author{Zheng-Yu Weng}
\affiliation{Institute for Advanced Study and Collaborative Innovation Center of Quantum Matter, Tsinghua University, Beijing, 100084, China}
\begin{abstract}

The Fermi sign structure plays a crucial role for a Landau's Fermi liquid. In this work, we identify the exact sign structure for the bipartite Hubbard model at an arbitrary strength of the on-site Coulomb repulsion $U$. This general sign structure naturally reproduces the conventional fermion signs in the small $U$ limit, and the phase string signs of the $t$-$J$ model in the large $U$ limit. We focus on the half-filling case as an example to illustrate why such a generic sign structure is important to understand the transition from the weakly correlated Fermi liquid regime to the strongly correlated Mott regime. In particular, we show that an electron fractionalization scheme with emergent partons and mutual Chern-Simons gauge fields provides a suitable framework to accurately handle the singular sign structure in two dimensions. A ground state ansatz at half-filling with incorporating the sign structure and the specific electron fractionalization is also proposed.

\end{abstract}
\date{\today}

\maketitle
%\tableofcontents

%%%%%%%%%%%%%%%%%%%%%%%%%%%%%%%%%%%%%%%%%%%%%%%%%%%%%%%%%%%%

\section{Introduction}

The celebrated Landau's Fermi liquid theory for weakly correlated electron systems has dominated the condensed matter physics for many decades. It is characterized by the low-energy quasiparticle excitations that can be adiabatically connected to the electron/hole excitations in a non-interacting Fermi gas, carrying the same conserved quantum numbers (charge, spin, and momentum). However, in a strongly correlated system, the Landau Fermi liquid theory is generally expected to break down \cite{Anderson1997}. 

A (doped) Mott insulator is a typical strongly correlated system, which is believed to be relevant to the high-$T_c$ cuprates \cite{Anderson1987}. Here a strong on-site Coulomb repulsion $U$ may drive the quasiparticles to fall apart into charge-neutral spinons and spinless chargons \cite{Lee2006}. Namely, as a non-Fermi-liquid state, the quasiparticle excitation may be replaced by collective excitations that carry quantum numbers distinct from the microscopic constituent particles -- the electrons. Concepts of \emph{electron fractionalization} and \emph{partons} are introduced to characterize non-Landau-quasiparticle excitations carrying a fraction of electron quantum numbers \cite{Anderson1997}.

Mathematically, such a fractionalization may be formally implemented by the so-called slave-particle scheme \cite{Lee2006} in an enlarged Hilbert space, where a local projection is usually required to eliminate the unphysical states of double occupancy to recover the physical Hilbert space. However, under the strict projection, this type of mathematical decomposition of the electron is \emph{not} unique and a statistical transmutation may generally take place among the partons. Since different fractionalizations will lead to inequivalent mean-field saddle-point states, which can only be connected through large (maybe uncontrollable) gauge fluctuations, a proper choice of the electron fractionalization is thus critical in finding the correct saddle-point for the true ground state.

Previously, based on the bipartite $t$-$J$ model as the large-$U$ Hubbard model, the sign structure has been rigorously shown to be completely changed from the conventional fermion signs of a non-interacting electron gas to much sparser phase string signs \cite{Wu2008}. In particular, at half-filling with the charge degree of freedom totally frozen, the whole statistical signs disappear, indicating that the altered statistical signs and the electron fractionalization are intimately connected (see. Sec. \ref{SecHAF}). 

In this work, for the first time we identify the general sign structure of the Hubbard model for an arbitrary strength of $U$. In the weak $U$ limit (as compared to the hopping integral $t$ of the tight-binding model), the conventional fermion signs of the free electrons will be recovered. In the opposite limit of large $U$, the aforementioned phase string signs will be reproduced. Such precise sign structure entangles the charge and spin degrees of freedom in a form of mutual statistics, which is expected to play an important role in the intermediate range of $U/t$.
 
In particular, we shall focus on the half-filling case, where the charge fluctuations influence the spin background increasingly with reducing $U/t$ via the quantum interference effect of the sign structure. Dictated by such sign structure, a new electron decomposition different from the conventional slave-particle schemes will be obtained. Using this specific fractionalization, a ground state ansatz can be constructed, which continuously interpolates between the weak and strong interaction limit at the half-filling. 

The remainder of the paper is organized as follows. In Sec. \ref{SecSS}, the partition function of the Hubbard model is expressed as the summation over the closed paths of spin and charge coordinates. The \emph{sign structure} will be identified, which determines whether a closed path contribution is constructive or destructive to the total partition function. It is an intrinsic property of the model independent of the choice of slave-particle formulations. In generally, the quantum interference of the singular sign structure cannot be treated perturbatively. In Sec. \ref{SecGSW} and Sec. \ref{SecEF}, we propose a ground state wavefunction ansatz at half-filling and an equivalent parton construction, respectively, in which the singular sign structure is naturally incorporated. In an analytic reformulation of the Hubbard model, the electron can be uniquely fractionalized into patrons in terms of the sign structure, whose effect is faithfully described by a pair of emergent mutual Chern-Simons gauge fields. The variational parameters that characterize the off-diagonal long range orders hidden in the parton subsystems can be determined either by a mean field theory or with a numerical variational study. Finally, Sec. \ref{SecDis} is devoted to summary and discussion. 
 
\section{Sign structure of the Hubbard model} \label{SecSS}

The Hubbard model is a minimal model widely adopted to describe correlated electrons on a lattice. It is given by
\begin{equation} \label{Hubbard}
H=-t\sum\limits_{\langle ij\rangle, \sigma} c_{i\sigma}^\dagger c_{j\sigma}+\mathrm{H.c.}+U\sum\limits_i n_{i\uparrow}n_{i\downarrow},
\end{equation}
where $c_{i\sigma}^\dagger$ ($c_{i\sigma}$) is the electron creation (annihilation) operator, $\mathrm{H.c.}$ stands for the Hermitian conjugate, and $n_{i\sigma}=c_{i\sigma}^\dagger c_{i\sigma}$ is the electron number operator. The $t$-term describes the electron hopping between the nearest neighboring sites, while the $U$-term introduces the repulsion between two electrons occupying the same site. 

The Hubbard model is expected to capture the rich phases in correlated electron systems. In the limit of $U/t\rightarrow 0$, it reduces to the tight-binding model of free electrons. In the opposite limit of $U/t\gg 1$, the double occupation is pushed out of the low energy Hilbert space. The sharp contrast can be most clearly illustrated at half-filling, where we have a weakly correlated itinerant electron state or a spin density wave (SDW) state due to the Fermi surface nesting at the small $U$ limit and a Mott insulator state at the large $U$ limit. How these two drastically distinct phases are connected in the intermediate correlation regime is currently subject to intensive investigations \cite{Hermele2007, Meng2010, He2012, Sorella2012, Hassan2013, Assaad2013a, Zhou2014, Chang2012b}.

Before addressing the general characterization of the Hubbard model, let us first focus on the simplest nontrivial case at half-filling in the large $U$ limit.

\subsection{Heisenberg model revisited} \label{SecHAF}

At half-filling and $U/t \gg 1$, each site is singly occupied such that the charge fluctuations are fully suppressed, leaving the localized spins as the only relevant degrees of freedom.
The virtual hopping process induces the Heisenberg superexchange coupling between the nearest neighboring spins
\begin{equation}\label{Hj}
H_J=J\sum\limits_{\langle ij\rangle }\vec{S}_i\cdot \vec{S}_j,
\end{equation}
with $J= 4t^2/U$.

Schwinger boson (SB) and Schwinger fermion (SF) are commonly adopted spin representations. In the SB (SF) formulation, $\vec{S}_i=\frac12 b_i^\dagger \vec{\sigma} b_i$, in which $b_i=(b_{i\uparrow},b_{i\downarrow})^{\mathrm{T}}$ are boson (fermion) operators and $\vec{\sigma}$ the Pauli matrices. Under the single-occupancy constraint $\sum_\sigma b_{i\sigma}^\dagger b_{i\sigma}=1$, both can rigorously reproduce the spin operator algebra.

But the mean field states obtained in the SB and the SF decompositions are very different. In fact, the ground state of $H_J$ is well described by the SB mean field theory \cite{Auerbach1988, Arovas1988, Yoshioka1989a, Sarker1989}. The SB mean field state on a bipartite lattice naturally possesses an antiferromagnetic long range order (AFLRO) \cite{Auerbach1988, Arovas1988, Yoshioka1989a, Sarker1989}, whereas the best SF mean field state is the $\pi$-flux state with an algebraic spin correlation \cite{Affleck1988, Marston1989}. In the latter case, gauge fluctuations beyond the mean field approximation need to be introduced to partially account for the local projection of the single-occupancy constraint, where the staggered spin correlation is enhanced to diverge \cite{Rantner2002}.

Why are the correlations in these two mean field states so distinct? Generally speaking, the projection of the single-occupancy constraint eliminates the unphysical states in both formulations. Then why does the projection seem important in the SF description, while it plays an insignificant role in the SB formulation?

On a bipartite lattice, we adopt the rotated Ising basis in the SB formulation \cite{Marshall1955}, $b_{i\sigma}\mapsto (-\sigma)^i b_{i\sigma}$, in which $(-\sigma)^i=1$ for $i\in A$ and $-\sigma$ for $i\in B$ sublattice. The Heisenberg model Eq. (\ref{Hj}) is then reexpressed as
\begin{equation} \label{EqSBHJ}
\begin{split}
H_J=&-\frac12 J\sum\limits_{\langle ij\rangle}(b_{i\uparrow}^\dagger b_{j\downarrow}^\dagger b_{i\downarrow}b_{j\uparrow}+b_{i\downarrow}^\dagger b_{j\uparrow}^\dagger b_{i\uparrow}b_{j\downarrow})+H_{\mathrm{dt}}\\
\equiv& H_{\uparrow\downarrow}+H_{\mathrm{dt}}
\end{split}
\end{equation}
where $H_{\mathrm{dt}}$ stands for the diagonal terms in the Ising basis, $\big\{|\alpha\rangle = b_{i_1\uparrow}^\dag\cdots b_{i_M\uparrow}^\dag b_{j_1 \downarrow}^\dag \cdots b_{j_{N-M} \downarrow}^\dag |0\rangle\big\}$ with $M$ denoting the total number of up-spins and $N$ the lattice size. Making a high temperature expansion to all orders and inserting the orthonormal basis \cite{Wu2008} $\sum_{\alpha}|\alpha\rangle\langle \alpha|$, we find
\begin{equation}\label{Eqexpansion}
\mathcal{Z}=\mathrm{tr}e^{-\beta H_J}=\sum\limits_{n=0}^\infty \frac{\beta^n}{n!}\sum\limits_{\alpha_i}\prod\limits_{i=0}^{n-1}\langle \alpha_{i+1}|(-H_J)|\alpha_i\rangle,
\end{equation}
where $|\alpha_n\rangle =|\alpha_0\rangle$. So we find all particles form closed paths and the partition function is cast into a summation over the closed paths $c$,
\begin{equation}\label{SBSignStructure}
\mathcal{Z}=\sum\limits_{c}W_J[c],\quad W_J[c]\geq 0.
\end{equation}
Contributions from all paths are positive because all the off-diagonal elements of the Hamiltonian Eq (\ref{EqSBHJ}) are negative while the diagonal terms do not affect the sign structure (see Appendix \ref{DiagonalTerms} for the rigorous proof).

In the SF formulation, we can similarly obtain the following formal expansion,
\begin{equation}\label{SFSignStructure}
\mathcal{Z}=\sum\limits_{\{c\}}(-1)^{P_\uparrow[c]+P_\downarrow[c]}W_J[c],\quad W_J[c]\geq 0,
\end{equation}
where $P_\sigma[c]$ is the parity of the spin-$\sigma$ permutation due to the \emph{fermion signs} and $W_J[c]$ is the same as in Eq. (\ref{SBSignStructure}).

With the no-double-occupancy constraint relaxed in the mean field theory, the sign structures in Eqs. (\ref{SBSignStructure}) and (\ref{SFSignStructure}) look apparently incompatible. The fermion signs in Eq. (\ref{SFSignStructure}) cause destructive interference among different paths and lead to the non-divergent spin structure factor in the mean field state. In the SB formulation, on the contrary, without the statistical signs, all closed paths of free bosons interfere constructively, which results in the divergence of the spin structure factor and the emergence of the AFLRO.

But under the strict single-occupancy constraint, the particle permutation is only realized by successive swaps and the parity of the total swap number equals to the permutation parity within each closed path, i.e., $(-1)^{S_{\uparrow\downarrow}[c]}=(-1)^{P_\uparrow[c]+P_\downarrow[c]}$, in which $S_{\uparrow\downarrow}[c]$ denotes the total number of swaps between an up- and a down-spin. In particular, on a bipartite lattice, $S_{\uparrow\downarrow}[c]\equiv 0\pmod{2}$, because the parity of spin-up particle number on a given sublattice changes at each swap, which is always restored after a closed path. Therefore, all closed paths contribute positively, thus the spinon Bose condensation picture in the SB formulation is justified and AFLRO ensues \cite{Yoshioka1989a, Sarker1989}.

As shown above, the Hilbert space restriction can lead to \emph{statistical transmutation}, i.e., the particle statistics inherent in the true ground state can be distinct from that in the original slave-particle formulation. Even though both SB and SF are mathematically equivalent under the strict enforcement of the constraint, they lead to physically quite different mean field states if the constraint is relaxed. Therefore, in order to capture the true ground state, a ``correct'' fractionalization formulation is important to start with. Since the statistics of the underlying particles are generally encoded in the sign structure, finding a proper fractionalization requires to precisely identify the \emph{ irreducible} sign structure in the partition function. 

\subsection{General sign structure of the Hubbard model} \label{SecHSS}

Now we derive the general sign structure of the Hubbard model in Eq. (\ref{Hubbard}) for an arbitrary $U$. Its physical consequences will be made clear in Sec. \ref{SecGSW}. In the following, we shall use the slave-fermion representation \cite{Yoshioka1989}, and in Appendix \ref{AppEquiv} we show that the same sign structure can be equivalently obtained in the slave-boson representation. In other words, the intrinsic sign structure is independent of the detailed mathematical formulation.

We may formally define the singly occupied sites as the \emph{ spinon} states and the sites of double-occupation and empty as \emph{doublons} and \emph{holons}, respectively, by making the following mapping
\begin{align}\label{mapping}
&c_{i\uparrow }^{\dagger }|0\rangle \mapsto (-1)^i b_{i\uparrow }^{\dagger }|0\rangle,\quad c_{i\downarrow}^{\dagger }|0\rangle \mapsto b_{i\downarrow}^{\dagger }|0\rangle ,\\
&c_{i\uparrow }^{\dagger }c_{i\downarrow }^{\dagger }|0\rangle \mapsto (-1)^id_i^{\dagger }|0\rangle,\quad |0\rangle \mapsto h_i^{\dagger }|0\rangle,
\end{align}
The corresponding electron operator can be reexpressed in the slave-fermion representation as follows
\begin{equation}\label{coperator}
c_{i \sigma }=(-\sigma)^i (h_i^{\dagger }b_{i \sigma }+\sigma b_{i -\sigma }^{\dagger } d_i).
\end{equation}
The electron $c$-operator algebra is realized by \emph{fermionic} chargons ($d_i$ and $h_i$) and \emph{bosonic} spinons ($b_{i\sigma}$) together with the following constraint on the physical Hilbert space
\begin{equation}\label{constraint}
\sum _{\sigma } b_{i \sigma }^{\dagger }b_{i \sigma }+d_i^{\dagger }d_i+h_i^{\dagger }h_i=1.
\end{equation}
It is noted that a staggered sign factor $(-1)^i$ in Eqs. (\ref{mapping}) and (\ref{coperator}) is explicitly introduced just for convenience in later counting of the total sign structure -- it reflects the Marshall sign rule for the Heisenberg model \cite{Marshall1955}.

\begin{figure}
\centering
\includegraphics[width=0.4\textwidth]{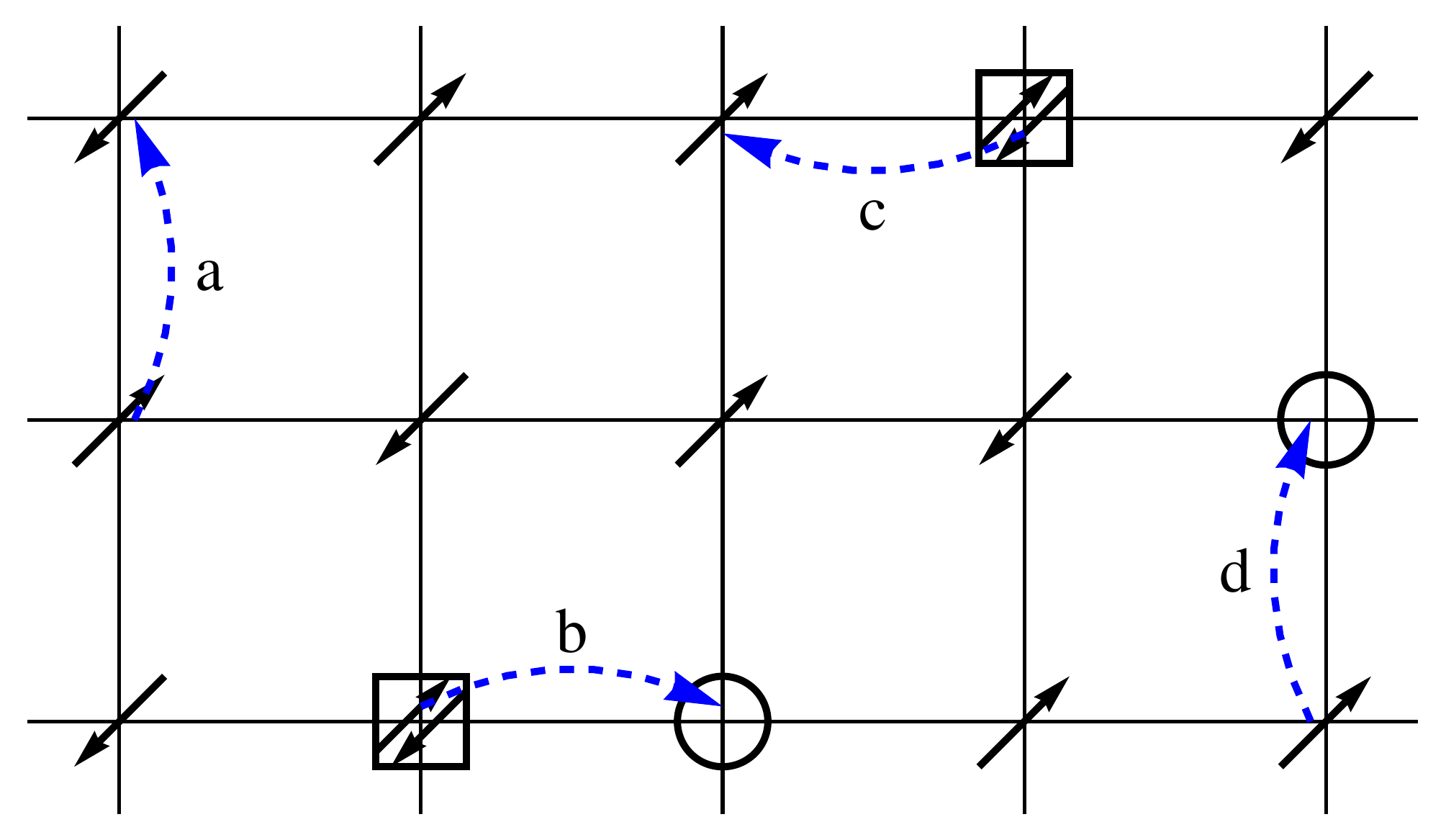}
\caption{(Color online) The elementary processes introduced by the kinetic term in Eq. (\ref{Ht}). The dashed arrows indicate the hopping of electrons. (a, b) $H_t$ creates or annihilates a chargon pair. (c, d) $H_t$ moves a doublon or a holon.}
\label{FigHt}
\end{figure}

Substituting Eq. (\ref{coperator}) into the Hamiltonian Eq. (\ref{Hubbard}), we find
\begin{equation}\label{Ht}
\begin{split}
H_t=&-t\sum _{\langle i j\rangle } \Big[\sum _{\sigma } \big(b_{i \sigma }^{\dagger }b_{j -\sigma }^{\dagger }h_id_j+ b_{j\sigma}^\dag b_{i-\sigma}^\dag h_j d_i \big)\\
&+\big(b_{i\uparrow }^{\dagger }b_{j\uparrow }d_j^{\dagger }d_i+b_{i\uparrow }^{\dagger }b_{j\uparrow }h_j^{\dagger }h_i\big)\\
&-\big(b_{i\downarrow }^{\dagger }b_{j\downarrow }d_j^{\dagger }d_i+b_{i\downarrow }^{\dagger }b_{j\downarrow }h_j^{\dagger }h_i\big)\Big]+\mathrm{H.c.}\\
\equiv& -t\sum _{\langle i j\rangle } \big(P_{i j}+E_{i j}^{\uparrow }-E_{i j}^{\downarrow }\big)+\mathrm{H.c.}
\end{split}
\end{equation}
where $P_{ij}$ creates a spinon pair and annihilates a chargon pair and $E_{ij}^\sigma$ swaps a chargon with a spinon with spin $\sigma$ at the nearest neighboring site (see Fig. \ref{FigHt}). The $U$-term is given by
\begin{equation}\label{Hu}
\begin{split}
H_U=&U\sum _i d_i^{\dagger }d_i\\
=&\frac{1}{2}U\sum _i \big(d_i^{\dagger }d_i+h_i^{\dagger }h_i \big)+\frac{1}{2}U\big(N_e-N \big)
\end{split}
\end{equation}
where $N_e$ denotes the total number of electrons and $N$ the number of lattice sites, with
\begin{equation}\label{Ne}
N_e-N=\sum _i \big(d_i^{\dagger }d_i-h_i^{\dagger }h_i \big).
\end{equation}
Here $H_U$ is diagonal, while $H_t$ is off-diagonal on the basis formed by
\begin{equation}\label{basis}
\Big\{d_{l_1}^{\dagger }\cdots h_{m_1}^{\dagger }\cdots b_{i_1\uparrow}^{\dagger }\cdots b_{j_1\downarrow}^{\dagger }\cdots|0 \rangle\Big\},
\end{equation}
in which the constraint Eq. (\ref{constraint}) is always satisfied.

We can then formally expand the partition function as outlined in Eq. (\ref{Eqexpansion}). A straightforward manipulation shows that the minus signs in the expansion come from the off-diagonal term $-t E_{ij}^\downarrow$ in $H_t$, in addition to the fermion signs of the holons and doublons; therefore, the partition function can be eventually written in a compact form
\begin{equation}\label{partition}
\mathcal{Z}=\sum _{c} (-1)^{N[c]}W_\text{H}[c],\quad W_\text{H}[c]\geq 0,
\end{equation}
with the sign factor given by
\begin{equation}\label{signfactors}
(-1)^{N[c]}= (-1)^{P_d[c]+P_h[c]+S_{d\downarrow }[c]+S_{h\downarrow }[c]},
\end{equation}
where $P_{d (h)}[c]$ denotes the parity of the permutation of doublons (holons) due to their fermion signs and $S_{d(h)\downarrow }[c]$ the total number of swaps of a doublon (holon) with a down-spinon due to the hopping term $-t E_{ij}^\downarrow$. Note that for a closed path, the number of exchanges between the holons and doublons are always parity even and do not contribute to the sign structure in Eq. (\ref{signfactors}).

Each closed particle path $c$ in Eq. (\ref{partition}) is weighted by a positive-definite weight
\begin{equation}
W_\text{H}[c]=(\beta t)^{N_t[c]} F_{N_t[c]}\big(-n_0\beta U,-n_1\beta U,\ldots, -n_{N_t[c]}\beta U\big),
\end{equation}
in which $N_t[c]$ is the total number of times that $H_t$ acts and $F_k$ is a multi-variable positive function defined in Appendix \ref{DiagonalTerms} which captures the contribution from the diagonal term $H_U$. $n_i$ ($i=0,1,\ldots, N_t[c]$) denotes the total number of doublons in the state $|\alpha_i\rangle$.

\subsubsection{Large $U/t$ limit}

\emph{Half filling}. At half-filling, we have $N_e=N$ and $\sum _i \big(d_i^{\dagger }d_i-h_i^{\dagger }h_i \big)=0$ [see Eq. (\ref{Ne})]. In the limit of $U/t\gg 1$, we further find $\sum_i \big(d_i^{\dagger }d_i+h_i^{\dagger }h_i\big)\rightarrow 0$ due to the large repulsion energy.

Then, in the partition function Eq. (\ref{partition}), those closed paths involving holons or doublons have vanishing weight $W_H[c]$. Consequently, for all the paths having non-vanishing contributions,
\begin{equation} \label{none}
N[c]= P_d[c]+P_h[c]+S_{d\downarrow }[c]+S_{h\downarrow }[c]=0,
\end{equation}
which is consistent with the result of the Heisenberg model as discussed in Sec. \ref{SecHAF}.

\emph{Finite doping}. Due to the particle-hole symmetry, we may only focus on the hole-doped side. In the limit of $U/t\gg 1$, the doublon number vanishes, $\sum_i d_i^{\dagger }d_i\rightarrow 0$ and the sign structure is reduced to
\begin{equation} \label{hole}
N[c]= P_h[c]+S_{h\downarrow }[c],
\end{equation}
which has been previous identified for the $t$-$J$ model \cite{Wu2008}. For the electron-doped case, we can simply change the subscript $h$ to $d$ in Eq. (\ref{hole}).

\emph{One hole case}. For the $t$-$J$ model with only one hole doped into the AF background, the sign structure is further reduced into
\begin{equation}
N[c]=S_{h\downarrow}[c],
\end{equation}
namely, the hole hopping leaves a string of signs (phase string) behind it, which is sensitive to the background spin configuration. In the presence of AFLRO, the spin flips cost little energy; consequently, the quasiparticle weight vanishes due to the destructive interference of different hopping paths \cite{Sheng1996, Weng1997} and the hole is predicted to be self-localized \citep{Weng2001}. Recently, the charge localization has been demonstrated for both single-hole-doped even and odd-leg ladders by large-scale density matrix renormalization group (DMRG) simulations \citep{Zhu2013}. In the DMRG study, the hole localization has been found to be true even in the spin gapped even-leg ladders, where the phase strings can still accumulate to cause quantum destructive interference over a sufficiently large distance.

\subsubsection{Small $U/t$ limit}

In the small $U/t$ limit, the conventional fermion signs of electrons are expected to be recovered. Here the energy cost for creating or annihilating a holon-doublon pair vanishes. With the proliferation of chargons, the original electron representation becomes more natural without redundancy.

\begin{figure}
 \centering
 \includegraphics[width=0.4\textwidth]{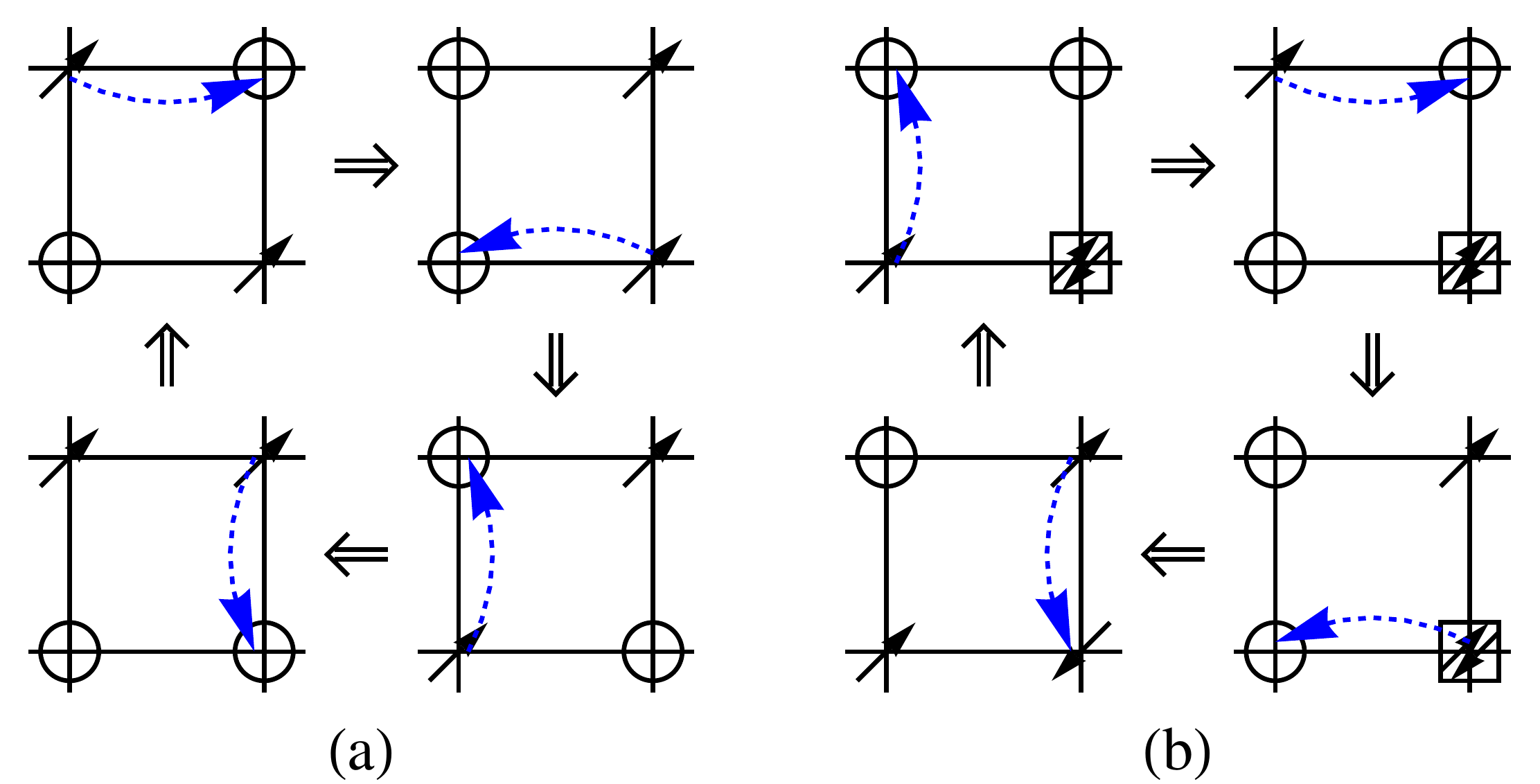}
 \caption{(Color online) (a) The exchange of two spin-$\uparrow$ electrons induces the exchange of two holes, thus $P_{\uparrow}^e[c]=P_{h}[c]=1$. (b) In terms of partons, the process involves the exchange of two holons, $P_h[c]=1$, while in terms of electrons, $P_\uparrow^e[c]=1$.}
 \label{FigPermute}
\end{figure}

Indeed the sign factor Eq. (\ref{signfactors}) can be reexpressed in terms of the electrons
\begin{equation}\label{fermisign}
(-1)^{N[c]}= (-1)^{P^e_{\uparrow}[c]+P^e_{\downarrow}[c]},
\end{equation}
where $P^e_{\sigma}[c]$ is the parity of the permutation between electrons of spin $\sigma$, representing the conventional fermion sign structure. The rigorous proof of Eq. (\ref{fermisign}) is given in Appendix \ref{AppEquiv}. For example, the exchange between two single-occupancy electrons (spinons) of spin-$\uparrow$ is illustrated in Fig. \ref{FigPermute} (a). On the other hand, the process shown in Fig. \ref{FigPermute} (b) involving the hopping of a spinon and the annihilation and creation of holon-doublon pairs can be naturally counted by the fermion sign of the two electrons, with the doublon expressed in terms of electron double occupancy. In both processes, the counts of the minus signs on both sides of Eq. (\ref{fermisign}) are equivalent.

Therefore, in the weak interaction limit, the fermion sign structure is naturally recovered in the electron representation as given in Eq. (\ref{fermisign}). But in the opposite limit of strong interaction, Eq. (\ref{hole}) becomes the irreducible sign structure. In particular, it can further reduce to be sign-free at half-filling. This is a drastic departure from the full fermion signs given in Eq. (\ref{fermisign}) that become maximally redundant here, and thus must be removed by strictly enforcing the no-double-occupancy constraint on the possible paths via the positive weight $W_H[c]$. In other words, if we adopt the electron representation to deal with the large $U/t$ Hubbard model, the unphysical sign structure will cause a fundamental difficulty unless we can handle the constraint rigorously. As discussed in Sec. \ref{SecHAF}, at half-filling, the distinction between the slave-boson and slave-fermion representations in the sign structure can be reconciled only under the strict constraint and the relaxation of the latter also results in drastically different mean field states. In general, how to properly formulate the electron fractionalization in the correlated regime at intermediate $U/t$ should be tied to how to correctly incorporate the irreducible sign structure given in Eq. (\ref{signfactors}), which we turn to in the following sections.

\section{Ground-State Wavefunction} \label{SecGSW}

In the above section, the precise sign structure Eq. (\ref{signfactors}) for the Hubbard model has been identified. One of the most prominent features is that such a sign structure will continuously deviate from the conventional fermion signs, in a direction by becoming ``sparser'' with the increase of $U/t$. In fact, the nontrivial statistical signs in Eq. (\ref{signfactors}) will wholly diminish at half-filling, once the charge (holon-doublon) fluctuations are suppressed by a large $U/t$. In the following, we explore how to generally incorporate this non-Fermi sign structure into a wavefunction formalism.

\subsection{Sign structure in ground-state wavefunction} \label{SecSSGSW}

Similar to the proof by Wu, Weng and Zaanen for the $t$-$J$ model case \cite{Wu2008}, the ground state wavefunction can be directly related to the sign structure as follows
\begin{equation}\label{GSCondition}
\langle \alpha |\Psi_{\text{G}}\rangle \langle \Psi_{\text{G}}|\alpha\rangle \propto\sum\limits_{c(\alpha,\alpha)}(-1)^{N[c(\alpha,\alpha)]}W[c(\alpha,\alpha)],\quad W[c(\alpha,\alpha)]\geq 0,
\end{equation}
where $|\alpha\rangle$ denotes an arbitrary spinon-chargon configuration and the ground state $|\Psi_{\text{G}}\rangle$ is assumed to be non-degenerate for simplicity. The summation is over all closed paths $c(\alpha,\alpha)$ that both start from and end with the state $|\alpha\rangle$ with a positive weight $W[c(\alpha,\alpha)]$.

For two arbitrary spinon-chargon configurations, $|\alpha\rangle$ and $|\alpha'\rangle$, we can make the high-temperature expansion to obtain
\begin{equation}\label{HTexpansion}
\begin{split}
\langle \alpha' |e^{-\beta H}|\alpha\rangle =&\sum\limits_{n=0}^\infty \frac{\beta^n}{n!}\sum\limits_{\alpha_i}\prod\limits_{i=0}^{n-1}\langle \alpha_{i+1}|(-H)|\alpha_i\rangle \\
=&\sum\limits_{c(\alpha,\alpha')}(-1)^{N[c(\alpha,\alpha')]}W[c(\alpha,\alpha')],
\end{split}
\end{equation}
where $W[c(\alpha,\alpha')]\geq 0$ and
\begin{equation}\label{EqSSgen}
(-1)^{N[c(\alpha,\alpha')]}\equiv \mathrm{sgn}(\langle \alpha_n|(-H_t)|\alpha_{n-1}\rangle \cdots \langle \alpha_1|(-H_t)|\alpha_0\rangle).
\end{equation}
with $|\alpha_0\rangle=|\alpha\rangle$ and $|\alpha_n\rangle=|\alpha'\rangle$. In the limit $\beta\rightarrow \infty$, the left-hand-side of Eq. (\ref{HTexpansion}) further reduces to
\begin{equation}\label{HTexpansion1}
\langle \alpha' |e^{-\beta H}|\alpha\rangle \rightarrow e^{-\beta E_{G}} \langle \alpha' |\Psi_{\text{G}}\rangle \langle \Psi_{\text{G}}|\alpha\rangle,
\end{equation}
so we find
\begin{equation}\label{GSCondition1}
 \langle \alpha' |\Psi_{\text{G}}\rangle \langle \Psi_{\text{G}}|\alpha\rangle \rightarrow e^{\beta E_{G}} \sum\limits_{c(\alpha,\alpha')}(-1)^{N[c(\alpha,\alpha')]}W[c(\alpha,\alpha')]
\end{equation}
with Eq. (\ref{GSCondition}) reproduced for $|\alpha'\rangle=|\alpha\rangle$.

Hence, the sign structure $(-1)^{N[c]}$ given in Eq. (\ref{signfactors}) explicitly appears in the ground state and plays a critical role through the singular destructive interference in the summation over all the possible paths. Such a sign structure in Eq. (\ref{GSCondition}) may be regarded as a generalized Berry's phase factor for particles in the state $|\alpha\rangle$ traversing through a closed path $c(\alpha,\alpha)$, which is composed of multiple spatial loops independent of time. Furthermore, $N[c]$ just counts the parity of the swaps between spinons and chargons, so the sign structure is not simply a path-dependent geometric Berry phase -- it is topological, indicating the emergent particle statistics. Such new statistical signs are singular, fluctuating between $+1$ and $-1$, which cannot be treated perturbatively and thus should be built into the wavefunction ansatz as a priori, similar to the conventional fermion signs in a Fermi liquid state.

We can similarly show that such singular signs also appear in the correlation functions \cite{Wu2008}. Because the formal high-temperature expansion in Eq. (\ref{Eqexpansion}) and the expansion over the close paths in Eq. (\ref{SBSignStructure}) are the starting point of the stochastic series expansion (SSE) simulation \cite{Sandvik1991}, in which the weight $W_{J}[c]$ of each close path is taken as the Monte Carlo (MC) sampling probability, the negative signs $(-1)^{N[c]}$ appearing in the expansion of the Hubbard model Eq. (\ref{partition}) prevent the interpretation of $(-1)^{N[c]}W_{\mathrm{H}}[c]$ as the MC probability, thus make the SSE simulation of the Hubbard model in general impossible, since the required MC time cost scales exponentially with the lattice size and the inverse temperature $\beta$ \cite{Troyer2005}.

\subsection{Ground state ansatz at half-filling} \label{SecGSA}

In the strong coupling limit $U\gg t$, the Hubbard model reduces to the Heisenberg model in Eq. (\ref{Hj}) at half-filling, and the ground state is well-described by the Liang-Doucot-Anderson type of bosonic resonating valence bond (RVB) state \cite{Liang1988} $|b\text{-RVB}\rangle$. Here $|b\text{-RVB}\rangle$ is a variational state, which can be expressed in terms of the $b$-spinons introduced in Sec. \ref{SecSS} as follows
\begin{equation}
|b\text{-RVB}\rangle= \hat{P}_s|\Phi _{b}\rangle ,
\label{lda}
\end{equation}
with
\begin{equation}
|\Phi _{b}\rangle = \exp \Bigg( \sum_{ij}W_{ij}b_{i\uparrow }^{\dagger
}b_{j\downarrow }^{\dagger }\Bigg) |0\rangle, \label{phirvb}
\end{equation}
in which the $b$-spinons are RVB-paired with an amplitude $W_{ij}$. The projection operator $\hat{P}_{s}$ enforces the single-occupancy constraint
$\sum_{\sigma} n_{i\sigma }^{b}=1$ ($n_{i\sigma }^{b}= b_{i\sigma }^{\dagger }b_{i\sigma }$) at half-filling. $|\Phi _{b}\rangle$ can be taken as a Schwinger boson mean field state, with $W_{ij}\propto |i-j|^{-3}$ for $i$ and $j$ belonging to different sublattices and 0 otherwise. The corresponding RVB state can describe the AF ground state accurately \cite {Liang1988}. $|b\text{-RVB} \rangle$ can also be expressed in terms of the electron operators as follows \cite{Weng2011},
\begin{equation}
|b\text{-RVB}\rangle =\sum_{\{ \sigma _{s}\}}\Phi _{\mathrm{RVB}}(\{\sigma_{s}\}) c_{1\sigma
_{1}}^{\dagger }c_{2\sigma _{2}}^{\dagger }\cdot \cdot \cdot c_{N\sigma
_{N}}^{\dagger }|0\rangle \label{RVB}
\end{equation}%
where $\Phi _{\mathrm{RVB}}\left( \{ \sigma _{s}\} \right) = \sum\prod_{(ij)}(-1)^{i}W_{ij}$ for each given spin configuration $\{ \sigma _{s}\}=\sigma _{1},\sigma _{2},\cdot \cdot\cdot ,\sigma _{N}$ and the summation is over all possible valence bond covering of the bipartite lattice.

Let us take $|b\text{-RVB}\rangle$ as our starting point. It does not involve any statistical signs because $b^{\dagger}_{i\sigma}$ is bosonic and $\Phi _{\mathrm{RVB}}$ is a bosonic wavefunction, in which the Marshall sign $(-1)^{i}$ does not contribute to the sign structure for closed paths. Previously, a ground state ansatz has been constructed in the large $U/t$ limit at finite doping based on the nontrivial sign structure in Eq. (\ref{hole}) accompanying the doped holes \cite{Weng2011}. In the following, we shall generalize this construction to finite $U/t$, where the nontrivial sign structure emerges even at half-filling when the charge fluctuations get involved.

Specifically, we propose the following ground state ansatz for the 2D Hubbard model at half-filling
\begin{equation}\label{WFAnsatz}
|\Psi_{\text{G}}\rangle =C e^{\hat{D}}|b\text{-RVB}\rangle,
\end{equation}
where $C$ is a normalization factor and $\hat{D}$ describes the charge fluctuations created as holon-doublon pairs, which maintain the charge neutrality at half-filling:
\begin{equation}\label{EqD}
\hat{D}=\sum\limits_{ij,\sigma}D_{ij}(\hat{c}_{i\sigma}^\dagger)_d(\hat{c}_{j\sigma})_h e^{i(\hat{\Omega}_i-\hat{\Omega}_j)} .
\end{equation}
Here the projected electron operators $(\hat{c}_{i\sigma}^\dagger)_d$ and $(\hat{c}_{j\sigma})_h$ create doubly occupied and empty sites, respectively, on the single-occupied spin background $|b\text{-RVB}\rangle$, i.e.,
\begin{align}
&(\hat{c}_{i\sigma}^\dag)_d |b\text{-RVB}\rangle =c_{i\sigma}^\dag |b\text{-RVB}\rangle,\label{Opd}\\
&(\hat{c}_{i\sigma}^\dag)_d |\text{non-single-occupancy}\rangle = 0;\label{Opd2}
\end{align}
and
\begin{align}
&(\hat{c}_{j\sigma})_h |b\text{-RVB}\rangle = c_{j\sigma} |b\text{-RVB}\rangle,\label{Oph}\\
&(\hat{c}_{j\sigma})_h |\text{non-single-occupancy}\rangle = 0.\label{Oph2}
\end{align}
So we find that $(\hat{c}_{i\sigma}^\dag)_d$ and $(\hat{c}_{j\sigma})_h$ are anticommutative: $\{(\hat{c}_{i\sigma}^\dag)_d, (\hat{c}_{j\sigma})_h\}=\{(\hat{c}_{i\sigma}^\dag)_d, (\hat{c}_{j\sigma})_d\}=\{(\hat{c}_{i\sigma}^\dag)_h, (\hat{c}_{j\sigma})_h\}=0$. The chargon permutation gives rise to the fermion signs $(-1)^{P_d[c]+P_h[c]}$.

The most singular part of the sign structure is captured by the phase shift factor $e^{i(\hat{\Omega}_i-\hat{\Omega}_j)}$ in Eq. (\ref{EqD}). Here the nonlocal phase shift is defined by
\begin{equation}\label{Omega}
\hat{\Omega}_i= \frac12 \sum\limits_{l\neq i}\theta_i(l)(n_{l\uparrow}^b-n_{l\downarrow}^b-1),
\end{equation}
where $\theta_i(l)=\mathrm{arg}(z_i-z_l)$ is the polar angle of the vector from site $l$ to site $i$. Because $|b\text{-RVB}\rangle$ is always at half-filling with $\sum_{\sigma}n_{l\sigma}^b=1$, we have $n^b_{i\uparrow} - n^b_{i\downarrow}-1 = -2n^b_{i\downarrow}$. As a chargon moves from site $i$ to $i'$, the phase factor $\hat{\Omega}_{i}$ changes by $-\sum_{l} n_{l\downarrow}^{b} (\theta_{i}(l)-\theta_{i'}(l))$, in which the down-spinon at site $l$ contributes $\theta_{i'}(l)-\theta_{i}(l)$, which equals to the angle swept by the vector from site $l$ to $i$ during the chargon motion. When the chargon completes a closed loop, a down-spinon in general contributes 0 (outside the loop) or $\pm 2\pi$ (inside the loop) phase shift, making no physical effect; however, a down-spinon lying on the loop, which exchanges with the chargon, contributes $\pm \pi$. So the phase string sign factor, $(-1)^{S_{d\downarrow}[c]+S_{h\downarrow}[c]}$, in Eq. (\ref{signfactors}) can be naturally realized by computing the Berry's phase contribution from $e^{i(\hat{\Omega}_i-\hat{\Omega}_j)}$. In this way, the ground state ansatz fully accommodates the sign structure.

The amplitude $D_{ij}$ in Eq. (\ref{EqD}) as well as $W_{ij}$ in Eq. (\ref{phirvb}) will be taken as variational parameters, which are presumably smooth since the singular sign structure has been incorporated in the phase shift factor in Eq. (\ref{EqD}).

In the large $U/t$ regime, the holon-doublon pairs are only virtual excitations and the chargons inside each pair are confined within a short distance such that the system is a Mott insulator. So the amplitude $D_{ij}$ decays exponentially with the distance $|i-j|$. We have $e^{\hat{D}}\sim 1$ and $|\Psi_{\text{G}}\rangle\sim |b\text{-RVB}\rangle$ in the long wavelength limit. The effect of the charge fluctuations on the spin-spin correlation in $|b\text{-RVB}\rangle$ is weak and perturbatively treatable. In particular, the AFLRO is expected to survive from the charge fluctuations at sufficiently large $U/t$.

At $U/t\rightarrow 0$, a large number of chargon pairs will be spontaneously created and annihilated without costing much energy. They overlap with each other such that the holon-doublon pairs dissolve into individual particles, with the pairing amplitude $D_{ij}$ reducing to $D_{ij}\sim g_i g_j$ as $|i-j|\rightarrow \infty$. So the chargons will constitute well-defined fermionic single-particle excitations in Eq. (\ref{WFAnsatz}). Self-consistently, the $b$-spinons in the background $|b\text{-RVB}\rangle$ will be forced to form short-range RVB and drop out of the low-energy spectrum. Then the phase shift operator $\hat{\Omega}_i$ gets cancelled out in a length scale larger than the RVB pair size, such that $e^{i\hat{\Omega}_i}\sim O(1)$. Consequently, the fermionic chargons propagate coherently and the system becomes itinerant. Note that in such an itinerant regime, the state allows a Fermi liquid description or an SDW ordered phase under different lattices as to be decided by $D_{ij}$.

\section{Electron fractionalization and emergent gauge fields} \label{SecEF}

In the ground state ansatz Eq. (\ref{WFAnsatz}), the spin AF correlation is characterized by the $|b\mathrm{-RVB}\rangle$ state, in which the spinons form RVB singlet from $A$ to $B$ sublattices. However, as a chargon hops from one site to another, a spinon is enforced to backflow to guarantee the single-occupancy condition, suggesting that the spin correlation is also affected by the chargon motion, which is not captured by the $b$-RVB state. On the other hand, to properly incorporate the sign structure, a nonlocal phase shift $\hat{\Omega}_i$ has been introduced into the ground state ansatz in Eqs. (\ref{WFAnsatz}) and (\ref{EqD}). For such a phase shift to be well-defined, it is crucial to distinguish the spin correlations involved in the RVB background and the chargon motion, respectively, in the ground state Eq. (\ref{WFAnsatz}). Therefore, we will introduce another species of backflow spinons, denoted by the creation operator $a^{\dagger}_{i\sigma}$, which reside on the same sites as chargons and hop with them, carrying the opposite spin quantum numbers to the $b$-spinons on the same sites to compensate them. As shown in Sec. \ref{SecGauge} and in Ref. \onlinecite{Weng2011} for the doped $t$-$J$ model case, the backflow spinons are turned into fermions by the dressed phase factors, so that the spinon RVB pairing and the backflow motion are explicitly distinguished. We would stress that this fractionalization scheme is required by the singular sign structure as a result of the electron correlation.

\subsection{Fractionalization description}

We may formally introduce the following decomposition to reexpress the projected electron operators defined in Eqs. (\ref{Opd})--(\ref{Oph2}) by
\begin{align}
(\hat{c}_{i\sigma}^\dagger)_d &=\hat{P} \sigma d_{i}^\dagger a_{i\sigma}^\dagger e^{-i\hat{\Omega}_i}(-\sigma)^{i}, \label{Opdf}\\
(\hat{c}_{i\sigma})_h &=\hat{P} h_{i}^\dag a_{i-\sigma}^\dag e^{i\hat{\Omega}_i}(-\sigma)^{i}, \label{Ophf}
\end{align}
which act on the single-occupancy state $|b\text{-RVB}\rangle$. Here $d_{i}^\dag$ ($h_{i}^{\dag}$) creates a \emph{bosonic} doublon (holon), and $a_{i\sigma}^\dag$ creates a \emph{fermionic} spinon carrying the opposite spin quantum number to compensate the underlying spin in $|b\text{-RVB}\rangle$ at the same site, known as the backflow spinon \cite{Weng2011}. The latter is enforced by the projection operator $\hat{P}$ (to be explicitly defined below) to realize the particle number constraints dictated by Eqs. (\ref{Opd2}) and (\ref{Oph2}). The sign factor $(-\sigma)^{i}$ in Eqs. (\ref{Opdf}) and (\ref{Ophf}) comes from the Marshall sign of the $|b\text{-RVB}\rangle$ state.

Then the ground state can be expressed as a projected direct product state of the fractionalized particles
\begin{equation}\label{WFFra}
\begin{split}
|\Psi_{\text{G}}\rangle = \hat{P}|\Phi_c\rangle \otimes |\Phi_a\rangle \otimes |\Phi_b\rangle,
\end{split}
\end{equation}
where the chargons are created and annihilated in $|\Phi_c\rangle$, forming holon-doublon pairs as follows
\begin{equation}\label{Phic}
\begin{split}
|\Phi_c\rangle =e^{\sum\limits_{ij}G_{ij}h_{i}^\dag d_{j}^\dag}|0\rangle
\end{split}
\end{equation}
and the backflow $a$-spinons are described by
\begin{equation}\label{Phia}
\begin{split}
|\Phi_a\rangle=e^{\sum\limits_{ij\sigma}g_{ij}a_{i\sigma}^\dag a_{j-\sigma}^\dag}|0\rangle
\end{split}
\end{equation}
with $D_{ij}=G_{ij}g_{ij}$. Finally the background $b$-spinons form a bosonic RVB state as described by $|\Phi_b\rangle$ in Eq. (\ref{phirvb}).

Here the projection operator $\hat{P}= \hat{P}_B\hat{P}_s$, in which $\hat{P}_s$ enforces the single-occupancy constraint, $\sum _{\sigma } n_{i\sigma}^b=1$ in the background state $|b\text{-RVB}\rangle= \hat{P}_s|\Phi _{b}\rangle$ as defined in Sec. \ref{SecGSA}, and $\hat{P}_B$ further implements the constraint 
\begin{align}\label{PB}
n_{i\sigma}^a=n_i^c n_{i-\sigma}^b, 
\end{align}
in which $n_{i\sigma}^a = a^{\dagger}_{i\sigma} a_{i\sigma}$, and $n_i^c=n_{i}^{h}+n^{d}_{i}$ denotes the total chargon number.

In this new fractionalization formulation, the singular sign structure is ``gauged away'' from the ground state Eq. (\ref{WFFra}) and the ground state is reduced to a direct product of three subsystems, each of which looks quite conventional. They are given by three sets of parameters, $G_{ij}$, $g_{ij}$, and $W_{ij}$ in Eq. (\ref{WFFra}), which can be determined by either a variational or a generalized mean field treatment.

In the following, we shall give a microscopic derivation of such a new fractionalization starting from the slave-fermion formalism. We will show that even though the singular sign structure is absorbed such that the phase shift factor $e^{i\hat{\Omega}_{i}}$ does not appear explicitly in the ground state Eq. (\ref{WFFra}), its physical effect of quantum interference cannot be gauged away and, as a matter of fact, will influence the three subsystems through emergent topological gauge fields.

\subsection{Emergent gauge fields} \label{SecGauge}

In the slave-fermion formulation Eq. (\ref{coperator}) of the Hubbard model, the origin of the main sign structure (i.e., the phase string effect) can be traced back to the minus sign in front of $E_{i j}^{\downarrow}$ in the hopping term Eq. (\ref{Ht}). 

Similar to the $t$-$J$ model case \cite{Weng1997}, we may introduce the following unitary transformation to explicitly incorporate the sign structure:
\begin{equation}
\hat{O}\mapsto e^{i \hat{\Theta }}\hat{O}e^{-i \hat{\Theta }},\quad\hat{\Theta }\equiv \sum _{i, l}^{i\neq l} (n_i^h-n_i^d)\theta _{i}(l)n_{l\downarrow}^b
\end{equation}
Then the Hamiltonian in the slave-fermion representation can be transformed into
\begin{equation}
\label{transfH}
\begin{split}
e^{i \hat{\Theta }}H_t e^{-i \hat{\Theta }}=& H_t^0+H_t^1,\\
e^{i \hat{\Theta }}H_U e^{-i \hat{\Theta }}=&U\sum_i d_i^\dagger d_i,
\end{split}
\end{equation}
where 
\begin{equation}
\label{Ht0}
H_t^0=-t\sum _{\langle i j\rangle} \left(\hat{\Delta}^s_{ij}\right)^{\dagger}\hat{\Delta}^c_{ij} +\mathrm{H.c.} 
\end{equation}
which involves the creation and annihilation of $b$-spinon and chargon pairs, with
\begin{align}
\hat{\Delta}^s_{ij}\equiv & \sum_{\sigma } e^{-i \sigma A_{ij}^c} b_{i \sigma }b_{j -\sigma },\label{Ds}\\
\hat{\Delta}^c_{ij}\equiv & e^{-i (A_{ij}^s-\phi _{ij}^0)} h_{i} d_{j}+e^{i(A^{s}_{ij}-\phi_{ij}^{0})}h_{j}d_{i}.
\label{Dc}
\end{align}
Here the link variables are defined by
\begin{align}
A_{i j}^c&= \frac{1}{2}\sum _{l\neq i,j} (\theta _{i}(l)-\theta _{j}(l))(n_{l}^h-n_{l}^d),\label{Acij}\\
A_{i j}^s&= \frac{1}{2}\sum
_{l\neq i,j} (\theta _{i}(l)-\theta _{j}(l))(n_{l\uparrow }^b-n_{l\downarrow }^b),\label{Asij}\\
\phi _{i,j}^0&= \frac{1}{2}\sum _{l \neq ij} (\theta _i(l)-\theta _j(l)). \label{phi0ij}
\end{align}
Note that the chargon operators are bosonized in Eqs. (\ref{Ht0}) and (\ref{Ht1}) by the following Jordan-Wigner transformations
\begin{align}
&h_ie^{-i\sum _{l\neq i} \theta _{l}(i)n_l^h}\mapsto h_{i},\quad e^{i\sum _{l\neq i} \theta _{l}(i)n_l^h}h_i^{\dagger
}\mapsto h_{i}^{\dagger },\\
&d_ie^{i\sum _{l\neq i} \theta _{l}(i)n_l^d}\mapsto d_{i},\quad e^{-i\sum _{l\neq i} \theta _{l}(i)n_l^d}d_i^{\dagger
}\mapsto d_{i}^{\dagger },
\end{align}
and the new chargon operators satisfy
\begin{align}
&[h_{i}, h_{j} ]=[h_{i}^{\dagger }, h_{j}^{\dagger }]=[h_{i}, h_{j}^{\dagger }]=0,\\
&[d_{i}, d_{j} ]=[d_{i}^{\dagger }, d_{j}^{\dagger }]=[d_{i}, d_{j}^{\dagger }]=0,\\
&\{h_{i,} d_{j}\}=\{h_{i}, d_{j}^{\dagger }\}=\{h_{i}^{\dagger },d_{j}\}=\{h_{i}^{\dagger },d_{j}^{\dagger }\}=0, \label{hd}
\end{align}
for $i\neq j$.

Another term in $H_t$ involves the hopping of chargons as given by
\begin{equation}
\label{Ht10}
\begin{split}
H_t^1=&-t\sum _{\langle i j\rangle} \Big(\sum_{\sigma}b_{i \sigma }^{\dagger }b_{j \sigma }e^{i \sigma A_{i j}^c} \Big)\\
&\times \Big(h_{j}^{\dagger}h_{i}e^{i (A_{ji}^s-\phi _{ji}^0)}+d_{j}^{\dagger}d_{i}e^{-i (A_{ji}^s-\phi _{ji}^0)}\Big) +\mathrm{H.c.}
\end{split}
\end{equation}
While $H_t^0$ is responsible for the superexchange interaction between the $b$-spinons after integrating out the charge fluctuations at large $U/t$, $H_t^1$ describes the process of chargon hopping. 

The key challenge in the large-$U$ Hubbard is how to properly treat the competing superexchange and hopping processes between $H_t^0$ and $H_t^1$ at finite doping. The similar competition will become increasingly important with reducing $U/t$ even at half-filling. Therefore, generalizing the same formulation in the $t$-$J$ model \cite{Weng2011}, we may expand the Hilbert space to introduce the $a$-spinon as follows
\begin{equation}\label{a}
a^{\dagger}_{i\sigma}e^{-i\sigma\sum_{l\neq i}\theta_i(l)n^a_{l\sigma}} \leftrightarrow b_{i-\sigma}(n_{i}^h+n_{i}^d)
\end{equation}
and meanwhile the chargon operators are mapped according to
\begin{align}
h_{i}\mapsto& h_{i}e^{-i\sum_{l\neq i}\theta_{i}(l)n_{l}^{d}\sum_{\sigma}\sigma n_{i\sigma}^{a}},\\
d_{i}\mapsto& d_{i}e^{i\sum_{l\neq i}\theta_{i}(l)n_{l}^{h}\sum_{\sigma}\sigma n_{i\sigma}^{a}},
\end{align}
so the anticommutation relations in Eq. (\ref{hd}) are turned into commutation relations and the chargon operators are fully bosonized. It is easy to check that $a_{i\sigma}^{\dagger}$ defined above is a fermionic operator, which corresponds to annihilating a $b$-spinon of spin $-\sigma$ on a site occupied by a chargon. In other words, one expands the Hibert space such that every site is always singly occupied by a $b$-spinon such that $\sum_{\sigma} n_{i\sigma}^b=1$. Then annihilating a $b$-spinon is equivalent to creating an $a$-spinon at the same site such that the total spin
\begin{equation}
{\bf S}_i={\bf S}^b_i+{\bf S}_i^a=0,
\end{equation}
with its spin ${\bf S}_i^a=-n_i^c{\bf S}^b_i$. The $b$-spin operators are expressed as
\begin{align}
S_{b}^{z}=&\frac12 \sum_{i,\sigma}\sigma b_{i\sigma}^{\dag}b_{i\sigma},\\
S^{+}_{b}=&\sum_{i}(-1)^{i}b_{i\uparrow}^{\dag}b_{i\downarrow}e^{i\sum_{l\neq i}\theta_{i}(l)(n_{l}^{h}-n_{l}^{d})}=(S^{-}_{b})^{\dag}
\end{align}
and the $a$-spin operators as
\begin{align}
S_{a}^{z}=&\frac{1}{2}\sum_{i,\sigma}\sigma a_{i\sigma}^{\dag}a_{i\sigma},\\
S^{+}_{a}=&\sum_{i}(-1)^{i}a_{i\uparrow}^{\dag}a_{i\downarrow}=(S^{-}_{a})^{\dag}.
\end{align}
The projected electron operator $\hat{c}_{i\sigma}$ which acts on the $|b\text{-RVB}\rangle$ background to create or annihilate the chargons and $a$-spinons is the combination of Eqs. (\ref{Opdf}) and (\ref{Ophf}),
\begin{equation}
\hat{c}_{i\sigma}=\hat{P}(h_{i}^{\dag}a_{i-\sigma}^{\dag}+\sigma d_{i}a_{i\sigma})e^{i\hat{\Omega}_{i}}(-\sigma)^{i}.
\end{equation}
It can be shown that the enlarged Hilbert space can reproduce the above physical Hilbert space by enforcing the constraint Eq. (\ref{PB}), which is precisely implemented by the projection $\hat{P}$ in the ground state ansatz Eq. (\ref{WFFra}).

Therefore, under the constraint Eq. (\ref{PB}), the hopping term $H_t^1$ in Eq. (\ref{Ht1}) may be further recast into the following form
\begin{align}\label{Ht1}
H_t^1=-t \sum\limits_{\langle ij\rangle} \sum\limits_{\sigma} a_{i\sigma}^\dag a_{j\sigma}\Big(h_{i}^\dag h_{j} e^{i(A_{ij}^s-\phi_{ij}^0)}+d_{i}^{\dag}d_{j}e^{-i(A_{ij}^{s}-\phi_{ij}^{0})}\Big) +\mathrm{H.c.},
\end{align}
where the $b$-spinon is replaced by the backflow $a$-spinon. Then $H_t^0$ and $H_t^1$ in Eqs. (\ref{Ht0}) and (\ref{Ht1}) together with $H_U$ in Eq. (\ref{Hu}), can serve as an appropriate exact starting point to study the variational ansatz state Eq. (\ref{WFFra}) at half-filling for arbitrary $U$ or even at arbitrary doping. The original singular sign structure hidden in the Hubbard model is now precisely incoporated by the link variables defined in Eqs. (\ref{Acij})--(\ref{phi0ij}).

Finally, the relation between the sign structure and the link variables (gauge fields) may be understood heuristically as follows. The sign factor $(-1)^{ S_{d(h)\downarrow}[c]}$ given by the total number of swaps of chargons with down-spinons can be interpreted as the Aharonov-Bohm phase due to a $-2\pi$ flux bound to each down-spinon (while none to up-spinons), which are seen by the chargons via $e^{i\hat{\Omega}_i}$. We set the $b$-spinons always half-filled, then the fluxes bound to spinons can be split into a static $-\pi$ flux on each site and a $\sigma \pi$ ($\sigma =\pm 1$) flux bound to each $\sigma$-spinon. The gauge potentials, denoted as $\phi_{ij}^0$ and $A_{ij}^s$, satisfy the following topological constraints on any spatial loop $C$:
\begin{align}
\sum\limits_{\langle ij\rangle \in C} \phi_{ij}^0&=\pi \sum\limits_{l~\text{inside}~C}1\bmod 2\pi,\label{EqTop1}\\
\sum\limits_{\langle ij\rangle \in C} A_{ij}^s&=\pi \sum\limits_{l~\text{inside}~C}(n_{l\uparrow}^b-n_{l\downarrow}^b)\bmod 2\pi.\label{EqTop2}
\end{align}
Doublons and holons carry $+1$ and $-1$ gauge charges of $A_{ij}^s$ respectively. These topological constraints characterize that the $A^s_{ij}$ and $\phi^0_{ij}$ fluxes are bound to spinons and plaquettes, respectively. On the other hand, making use of the charge-vortex duality, we find that chargons are also seen by $b$-spinons as $\pm \pi$ fluxes, i.e., there is another gauge potential $A_{ij}^c$ that couples to $b$-spinons and satisfies
\begin{equation}
\sum\limits_{\langle ij\rangle \in C} A_{ij}^c=\pi \sum\limits_{l~\text{inside}~C}(n_{i}^h-n_{i}^d)\bmod 2\pi,
\end{equation}
Thus, spinons and chargons are \emph{mutual semions} \cite{Weng1997}. The statistical interaction between them can be described by the mutual Chern-Simons gauge coupling between the gauge fields $A_{ij}^s$ and $A_{ij}^c$ \cite{Kou2005, Ye2011, Ye2012c}.

\section{Discussion} \label{SecDis}

The general sign structure for the Hubbard model has been rigorously identified in Eq. (\ref{signfactors}). It appears in the partition function of Eq. (\ref{partition}) through the summation over the closed paths known as the phase string effect \cite{Sheng1996, Weng1997}, which is further weighted by a positive amplitude $W_H$. Its very singular form will strongly affect the electron system by quantum interference effect. 

In the limit of vanishing $U/t$, such a sign structure reduces to that of non-interacting electrons, i.e., the well-known Fermi sign structure, and the consequence of quantum interference effect leads to a Fermi sea filled up by the electrons in the momentum space. In the large $U/t$ limit, however, the sign structure factor in Eq. (\ref{signfactors}) reduces to the trivial unity at half-filling due to the suppression of the charge fluctuations as controlled by $W_H$. Consequently the AFLRO is recovered as in the Heisenberg model, which is free from any destructive quantum interference of minus fermion signs. 

Therefore, in the intermediate coupling regime of $U/t$, the sign structure is expected to play an essential role to determine the Mott transition between a Fermi liquid/SDW state dictated by the fermion signs and a sign-free Mott insulator with the gapped charge (holon-doublon) fluctuations.

In order to provide a suitable starting point to study the nontrivial intermediate coupling regime, a ground state wavefunction is constructed in this work in Eq. (\ref{WFAnsatz}), which explicitly incorporates the sign structure and at the same time naturally interpolates between both weak and strong limits. Here the the phase string effect of the sign structure is encapsulated in the phase factor $e^{i\hat{\Omega}_{i}}$ in Eq. (\ref{Omega}). The latter then dictates a specific fractionalization scheme involving the background $b$-spinons, the chargons and the backflow $a$-spinons as shown in Eqs. (\ref{Opdf}) and (\ref{Ophf}), by which $e^{i\hat{\Omega}_{i}}$ is absorbed such that the ground state wavefunction can be cast into a projected direct product state of three subsystems in Eq. (\ref{WFFra}), each of them quite conventional. The variational coefficients in the ansatz state Eq. (\ref{WFFra}) can be determined by the unitary-transformed Hamiltonian Eqs. (\ref{Ht0}) and (\ref{Ht1}) in the electron fractionalization scheme. In particular, the quantum interference effect of the sign structure encoded in $e^{i\hat{\Omega}_{i}}$, which cannot be gauged away, now appears as a pair of gauge fields, $A_{ij}^{s}$ and $A_{ij}^{c}$, coupling to the chargons and $b$-spinons, respectively, to capture the mutual semion statistics between them.

Although the sign structure Eq. (\ref{signfactors}) is independent of the interaction strength and temperature, we note that at high temperatures, as more and more spinons and chargons are thermally excited, the propagations of both degrees of freedom are randomly scattered and severely decohered by the destructive interference due to the sign structure. Thus, the charge and spin transports in the high-temperature regime are expected to be highly incoherent (diffusive). On the other hand, at low temperature and especially in the ground state, the emergent partons can organize themselves according to the sign structure to reduce the quantum destructive interference and optimize the total energy. In the AFLRO phase, chargons are confined to form localized pairs on a long-range RVB background, where spin correlations are free from the charge frustrations. In the weak coupling regime, the $b$-RVB background becomes short-ranged and confined, where the electron quasiparticles, as the bound states of chargons and $a$-spinons, can propagate coherently. In the moderate coupling regime, where both $b$-spinons and chargons become quantum activated, the mutual semion statistics may render both of them phase incoherent. Therefore, exotic quantum disordered states (spin liquids) may set in and the emergent gauge theory presented in Sec. \ref{SecEF} can provide a qualitative analysis on the low energy physics where the fermionic $a$-spinons may account for the important low-energy spin excitations in the weak Mott insulator regime. A global quantum phase diagram with various exotic phases on different lattices at half-filling will be presented in a separate work.

\begin{acknowledgments}

We are grateful to A. Muramatsu, Q.-R. Wang, K. Wu, Y.-Z. You, and J. Zaanen for valuable discussions. This work was supported by the NBRPC Grant No. 2010CB923003.

\end{acknowledgments}

\begin{widetext}

\appendix

\section{Rigorous proof concerning the diagonal term} \label{DiagonalTerms}

In this appendix, we shall show that a diagonal term (denoted as $H_{\text{dt}}$) does not affect the sign structure. The particle configurations on both sides of a diagonal term are the same, so we can reorganize the expansion Eq. (\ref{Eqexpansion}) into a summation over particle paths with an indefinite number of $H_\text{dt}$ inserted at each step,
\begin{equation}\label{rigorous}
\mathcal{Z}=\sum_{\alpha_i}\beta^N\langle \alpha_{N}|\hat{O}_N|\alpha_{N-1}\rangle\langle \alpha_{N-1}|\hat{O}_{N-1}|\alpha_{N-2}\rangle\cdots\langle \alpha_1|\hat{O}_1|\alpha_{0}\rangle \bigg(\prod_{i=0}^{N}\sum_{k_i=0}^\infty\bigg)\frac{\beta^{\sum k_i}}{(N+\sum k_i)!}\prod_{i=0}^{N}\langle\alpha_{i}|(-H_\text{dt})|\alpha_i\rangle^{k_i}
\end{equation}
where $|\alpha_N\rangle =|\alpha_0\rangle$ and $\hat{O}_i$'s denote the off-diagonal terms. We shall show that the last factor in Eq. (\ref{rigorous}) is always positive for any given $N$ and $x_i\equiv \langle\alpha_{i}|(-H_\text{dt})|\alpha_i\rangle\in \mathbb{R}$.

Denote the multi-variable function
\begin{equation}
F_{N}(x_0,x_1,\ldots,x_N)\equiv \bigg(\prod\limits_{i=0}^N \sum\limits_{k_i=0}^\infty\bigg)\frac1{(N+\sum_i k_i)!}\prod\limits_{i=0}^N x_i^{k_i}.
\end{equation}
It is easy to see that if all $x_i> 0$,
\begin{equation}\label{EqFNPositive}
F_{N}(x_0,x_1,\ldots,x_N)>0.
\end{equation}

For $N=0$, $F_0(x_0)=\sum_{k_0=0}^\infty \frac{x_0^{k_0}} {k_0!}=e^{x_0}>0.$ For a general $N>0$, denoting $s\equiv \sum_{i=0}^N k_i$, we establish the relation between $F_N$ and $F_{N-1}$ as follows:

If $x_{N-1}\neq x_N$,
\begin{equation}
\begin{split}
F_N(x_0,x_1,\ldots,x_N)&=\sum\limits_{s=0}^\infty \frac{1}{(s+N)!}\sum\limits_{k_0=0}^{s}\sum\limits_{k_1=0}^{s-k_0}\cdots \sum\limits_{k_{N-1}=0}^{s-k_0-k_1-\cdots -k_{N-2}}x_0^{k_0}x_1^{k_1}\cdots x_{N-1}^{k_{N-1}}x_N^{s-k_0-\cdots -k_{N-1}}\\
&=\sum\limits_{s=0}^\infty \frac{1}{(s+N)!}\sum\limits_{k_0=0}^{s}\sum\limits_{k_1=0}^{s-k_0}\cdots \sum\limits_{k_{N-2}=0}^{s-k_0-k_1-\cdots -k_{N-3}}x_0^{k_0}x_1^{k_1}\cdots x_{N-2}^{k_{N-2}}\frac{x_{N-1}^{s-k_0-\cdots -k_{N-2}+1}-x_{N}^{s-k_0-\cdots -k_{N-2}+1}}{x_{N-1}-x_N}\\
&=\sum\limits_{s'=0}^\infty \frac{1}{(s'+N-1)!}\sum\limits_{k_0=0}^{s'}\sum\limits_{k_1=0}^{s'-k_0}\cdots \sum\limits_{k_{N-2}}^{s'-k_0-\cdots -k_{N-3}}x_0^{k_0} x_1^{k_1}\cdots x_{N-2}^{k_{N-2}}\frac{x_{N-1}^{s'-k_0-\cdots -k_{N-2}}-x_N^{s'-k_0-\cdots -k_{N-2}}}{x_{N-1}-x_N}\\
&=\frac{F_{N-1}(x_0,x_1,\cdots, x_{N-2},x_{N-1})-F_{N-1}(x_0,x_1,\cdots,x_{N-2},x_{N})}{x_{N-1}-x_N}.
\end{split}
\end{equation}
In the third line above, we defined $s'=s+1$. Define the ``difference ratio'' of a given function $f(x)$ as $d_x f(x)\big|_{x_1}^{x_2}\equiv \frac{f(x_1)-f(x_2)}{x_1-x_2}$, we find
\begin{equation}\label{EqFN}
F_N(x_0,x_1,\ldots,x_{N-2},x_{N-1},x_N)=d_x F_{N-1}(x_0,x_1,\ldots,x_{N-2},x)\big|_{x_{N-1}}^{x_N}.
\end{equation}

If $x_{N-1}=x_N$, using
\begin{equation}
\sum\limits_{k_{N-1}=0}^{s-k_0-k_1-\cdots -k_{N-2}}x_{N-1}^{k_{N-1}}x_N^{s-k_0-\cdots -k_{N-1}}=(s-k_0-\cdots -k_{N-2})x_{N-1}^{s-k_0-\cdots k_{N-2}}=\frac{\partial}{\partial x}x_{N-1}^{s-k_0-\cdots -k_{N-2}+1}\bigg|_{x=x_{N-1}},
\end{equation}
we find
\begin{equation}\label{EqFN2}
\begin{split}
F_{N}(x_0,x_1,\ldots, x_{N-1},x_N)&=\frac{\partial}{\partial x}\sum\limits_{s=0}^\infty \frac{1}{(s+N)!}\sum\limits_{k_0=0}^{s} \sum\limits_{k_1=0}^{s-k_0}\cdots \sum\limits_{k_{N-2}=0}^{s-k_0-\cdots -k_{N-3}}x_0^{k_0}x_1^{k_1}\cdots x_{N-2}^{k_{N-2}}x^{s-k_0-\cdots -k_{N-2}+1}\bigg|_{x=x_{N-1}}\\
&=\frac{\partial}{\partial x}\sum\limits_{s'=0}^\infty \frac{1}{(s'+N-1)!}\sum\limits_{k_0=0}^{s'} \sum\limits_{k_1=0}^{s'-k_0}\cdots \sum\limits_{k_{N-2}=0}^{s'-k_0-\cdots -k_{N-3}}x_0^{k_0}x_1^{k_1}\cdots x_{N-2}^{k_{N-2}}x^{s'-k_0-\cdots -k_{N-2}}\bigg|_{x=x_{N-1}}\\
&=\frac{\partial}{\partial x} F_{N-1}(x_0,x_1,\ldots, x_{N-2},x)\bigg|_{x=x_{N-1}},
\end{split}
\end{equation}
as expected from Eq. (\ref{EqFN}) by taking the limit $x_{N}\rightarrow x_{N-1}$.

Applying Eqs. (\ref{EqFN}) and (\ref{EqFN2}) iteratively, we find that $F_{N}(x_0,x_1,\ldots,x_N)$ is the $N$th order difference ratio of $e^x$.

From Eq. (\ref{EqFNPositive}), we know that if all $x_i>0$, the $N$th order difference ratio of $e^x$ is positive. If the general cases, taking $M>0$ such that all $x_i+M>0$, we find
\begin{equation}
F_{N}(x_0,\ldots,x_N)=e^{-M}F_N(x_0+M,\ldots,x_N+M)>0.
\end{equation}

\section{Relations of swap and permutation parities}\label{RelationPermutation}

On a lattice partons can only move by exchanging with each other, thus the parities of the total swap number and the permutation in a closed path $c$ are equal,
\begin{equation}
(-1)^{S[c]}=(-1)^{P[c]}.
\end{equation}
Several examples relevant to our work are listed below.
\begin{itemize}
 \item Heisenberg model: Swaps occur between an up-spinon and a down-spinon, $S[c]=S_{\uparrow\downarrow}[c]$. The permutation decomposes into that of the up-spinons and down-spinons, $(-1)^{P[c]}=(-1)^{P_{\uparrow}[c]+P_{\downarrow}[c]}$. Therefore,
 \begin{equation}
 (-1)^{S_{\uparrow\downarrow}[c]}=(-1)^{P_{\uparrow}[c]+P_{\downarrow}[c]}.
 \end{equation}
 which proves the equivalence of the sign structures of SB and SF formulations.
 \item If holons are introduced, e.g., in the $t$-$J$ model for doped Mott insulators, the swap processes can also take place between a holon and a spinon. We find
 \begin{equation}\label{EqRelationtJ}
 (-1)^{S_{\uparrow\downarrow}[c]+S_{h\uparrow}[c]+S_{h\downarrow}[c]}=(-1)^{P_\uparrow[c]+P_\downarrow[c]+P_h[c]}.
 \end{equation}
 \item In Hubbard model, both holons and doublons are present, but, a holon and a doublon cannot exchange with each other directly, i.e., $S_{hd}[c]=0$. Without invoking the chargon pair creation/annihilation processes, we find
 \begin{equation}\label{EqRelationHWO}
 (-1)^{S_{\uparrow\downarrow}[c]+S_{d\uparrow}[c]+S_{d\downarrow}[c]+S_{h\uparrow}[c]+S_{h\downarrow}[c]}
 =(-1)^{P_\uparrow[c]+P_\downarrow[c]+P_d[c]+P_h[c]}.
 \end{equation}
 In the presence of chargon pair creation/annihilation processes, the relation is modified due to the following type of processes
 \begin{equation}
 \uparrow\quad \downarrow \quad\underrightarrow{\hat{C}_{d\uparrow}} \quad\uparrow\downarrow\quad \_\_\quad \underrightarrow{\hat{A}_{d\downarrow}}\quad \downarrow\quad \uparrow
 \end{equation}
 that induce an effective swap between a pair of up- and down-spinons, where $\hat{C}_{d\sigma}$ ($\hat{A}_{d\sigma}$) is a chargon pair creation (annihilation) process with the doublon ($\sigma$-spinon) occupying the $\sigma$-spinon (doublon) site, therefor the spinon swap number parity $(-1)^{S_{\uparrow\downarrow}}$ in Eq. (\ref{EqRelationHWO}) should be replaced by $(-1)^{S_{\uparrow\downarrow}+(C_{d\uparrow}-A_{d\uparrow})}$ ($C_{d\sigma}$ and $A_{d\sigma}$ denote the numbers of $\hat{C}_{d\sigma}$ and $\hat{A}_{d\sigma}$ actions respectively) and we find
 \begin{equation}\label{EqRelationHW}
 (-1)^{S_{\uparrow\downarrow}[c]+C_{d\uparrow}[c]+A_{d\uparrow}[c]+S_{d\uparrow}[c]+S_{d\downarrow}[c]+S_{h\uparrow}[c]+S_{h\downarrow}[c]}
 =(-1)^{P_\uparrow[c]+P_\downarrow[c]+P_d[c]+P_h[c]}.
 \end{equation}
\end{itemize}

\subsection{Further simplification on a bipartite lattice}

On a bipartite lattice, each particle swap changes the particle number parities of the involved species on a sublattice, so in a \emph{closed} path, the total number of swaps involving a given species is even. We refer such constraints as ``parity rules'' below.

\begin{itemize}
 \item In the Heisenberg model, the parity rule of each spinon species requires that
 \begin{equation}
 (-1)^{S_{\uparrow \downarrow}[c]}=1.
 \end{equation}
 Therefore, the Heisenberg model on a bipartite lattice is free of sign problem as discussed in Sec. \ref{SecHAF}.
 \item In the $t$-$J$ model, the parity rules of holons and each species of spinons lead to
 \begin{equation}
 (-1)^{S_{h\uparrow}[c]+S_{h\downarrow}[c]}=(-1)^{S_{\uparrow\downarrow}[c]+S_{h\uparrow}[c]}
 =(-1)^{S_{\uparrow\downarrow}[c]+S_{h\downarrow}[c]}=1.
 \end{equation}
 We find
 \begin{equation}\label{EqRelationtJ2}
 (-1)^{S_{h\uparrow}[c]}=(-1)^{S_{h\downarrow}[c]}=(-1)^{S_{\uparrow\downarrow}[c]}.
 \end{equation}
 In the slave-boson formulation, $c_{i\sigma}=h_i^\dag f_{i\sigma}$, where $h_{i}^\dag$ and $f_{i\sigma}^\dag$ create bosonic holons and fermionic spinons respectively. The Hamiltonian is given by
 \begin{align}
 H_t&=-t\sum\limits_{\langle ij \rangle,\sigma} f_{i\sigma}^\dag h_j^\dag h_i f_{j\sigma}+\mathrm{H.c.}\\
 H_J&=-J\sum\limits_{\langle ij\rangle }f_{i\uparrow}^\dag f_{j\downarrow}^\dag f_{i\downarrow} f_{j\uparrow}+\mathrm{d.t.}
 \end{align}
 The sign structure is $(-1)^{P_\uparrow[c]+P_\downarrow[c]}$. Using Eqs. (\ref{EqRelationtJ},) and (\ref{EqRelationtJ2}), we find the equivalence of the sign structures of the slave-boson and slave-fermion \cite{Wu2008} formulations,
 \begin{equation}
 (-1)^{P_\uparrow[c]+P_\downarrow[c]}=(-1)^{P_h[c]+S_{h\downarrow}[c]}.
 \end{equation}
 \item The relationship of different parities for the Hubbard model and the equivalence of the sign structures in various representations are discussed in detail in App. \ref{AppEquiv}.
 \end{itemize}

\section{Equivalence of the sign structure of the Hubbard model} \label{AppEquiv}

In the Hubbard model, the creation and annihilation of chargon pairs also change the particle number parity. Denote $C_s^a$ ($A_s^a$) the total number of creation (annihilation) actions of particle species $s$ ($s=d,h,\uparrow,\downarrow$) on sublattice $a$ ($a=A,B$). The parity rules on sublattice $A$ lead to
\begin{equation}
\begin{split}
&(-1)^{S_{h\uparrow}[c]+S_{h\downarrow}[c]+C_h^A[c]+A_h^A[c]}=(-1)^{S_{d\uparrow}[c]+S_{d\downarrow}[c]+C_d^A[c]+A_d^A[c]}\\
=&(-1)^{S_{\uparrow \downarrow}[c]+S_{d\uparrow}[c]+S_{h\uparrow}[c]+C_{\uparrow}^A[c]+A_\uparrow^A[c]}=(-1)^{S_{\uparrow\downarrow}[c]+S_{d\downarrow}[c]+S_{h\downarrow}[c]+ C_{\downarrow}^A[c]+A_\downarrow^A[c]}=1.
\end{split}
\end{equation}
It is easy to see that the following relations hold in a closed path $c$,
\begin{equation}
\begin{split}
&C_d^A[c]=C_h^B[c],\quad C_d^B[c]=C_h^A[c],\quad A_d^A[c]=A_h^B[c],\quad A_d^B[c]=A_h^A[c],\\
&C_\uparrow^A[c]=C_\downarrow^B[c],\quad C_\uparrow^B[c]=C_\downarrow^A[c],\quad A_\uparrow^A[c]=A_\downarrow^B[c],\quad A_\uparrow^B[c]=A_\downarrow^A[c],
\end{split}
\end{equation}
and
\begin{equation}
\begin{split}
&C_d^A[c]+C_d^B[c]=C_h^A[c]+C_h^B[c]=C_\uparrow^A[c]+C_\uparrow^B[c]=C_\downarrow^A[c]+C_\downarrow^B[c]\\
=&A_d^A[c]+A_d^B[c]=A_h^A[c]+A_h^B[c]=A_\uparrow^A[c]+A_\uparrow^B[c]=A_\downarrow^A[c]+A_\downarrow^B[c],
\end{split}
\end{equation}
We have
\begin{equation}\label{EqHMB}
(-1)^{S_{d\uparrow}[c]+S_{d\downarrow}[c]+S_{h\uparrow}[c]+S_{h\downarrow}[c]}=(-1)^{C_d^A[c]+A_d^A[c]+C_h^A[c]+A_h^A[c]}=+1.
\end{equation}

\subsection{Slave fermion and slave boson in the rotated Ising basis}

On a bipartite lattice, the slave boson representation with the rotated Ising basis $c_{i\sigma}=(-\sigma)^i (h_i^\dag f_{i\sigma}+\sigma f_{i-\sigma}^\dag d_i)$ leads to
\begin{equation}
\begin{split}
 H_t=&-t\sum\limits_{\langle ij\rangle}(-f_{i\uparrow}^\dag h_j^\dag h_i f_{j\uparrow}+f_{i\downarrow}^\dag h_j^\dag h_i f_{j\downarrow}+d_i^\dag f_{j\downarrow}^\dag f_{i\downarrow}d_j-d_i^\dag f_{j\uparrow}^\dag f_{i\uparrow}d_j\\
 &+d_i^\dag h_j^\dag f_{i\downarrow}f_{j\uparrow}-d_i^\dag h_j f_{j\downarrow}f_{i\uparrow}-f_{j\uparrow}^\dag f_{i\downarrow}^\dag h_i d_j+f_{i\uparrow}^\dag f_{j\downarrow}^\dag h_i d_j)+\mathrm{H.c.}
\end{split}
\end{equation}
If we include a Heisenberg term, swaps between up- and down-spinons are also allowed,
\begin{equation}
 H_J=J\sum\limits_{\langle ij\rangle}\vec{S}_i\cdot \vec{S}_j
 = J\sum\limits_{\langle ij\rangle}f_{i\downarrow}^\dag f_{j\uparrow}^\dag f_{i\uparrow}f_{j\downarrow}+\mathrm{d.t.}
\end{equation}
The sign structure in the partition function is
\begin{equation} \label{EqSSSB}
 (-1)^{S_{d\uparrow}[c]+S_{h\uparrow}[c]+C_{d\uparrow}[c]+A_{d\uparrow}[c]+S_{\uparrow\downarrow}[c]+P_\uparrow[c]+P_\downarrow[c]}.
\end{equation}
Using Eqs. (\ref{EqRelationHW}) and (\ref{EqHMB}), we can show its equivalence to the slave fermion sign structure Eq. (\ref{signfactors}),
\begin{equation}
 (-1)^{S_{d\uparrow}[c]+S_{h\uparrow}[c]+C_{d\uparrow}[c]+A_{d\uparrow}[c]+S_{\uparrow\downarrow}[c]+P_\uparrow[c]+P_\downarrow[c]}
 =(-1)^{S_{d\downarrow}[c]+S_{h\downarrow}[c]+P_h[c]+P_d[c]}.
\end{equation}

\subsection{Slave boson and electron representations}

In terms of the slave boson representation,
\begin{equation}
 c_{i\sigma}=h_i^\dag f_{i\sigma}+\sigma f_{i-\sigma}^\dag d_i,
\end{equation}
the hopping term is given by
\begin{equation}
 \begin{split}
 H_t=&-t\sum\limits_{\langle ij\rangle} \bigg(\sum\limits_\sigma f_{i\sigma} ^\dag h_j^\dag h_i f_{j\sigma}-\sum\limits_\sigma f_{j\sigma}^\dag d_i^\dag d_jf_{i\sigma}+d_i^\dag h_j^\dag f_{j\uparrow}f_{i\downarrow} +d_i^\dag h_j^\dag f_{i\uparrow}f_{j\downarrow}+f_{j\downarrow}^\dag f_{i\uparrow}^\dag h_id_j +f_{i\downarrow}^\dag f_{j\uparrow}^\dag h_i d_j\bigg) +\mathrm{H.c.}
 \end{split}
\end{equation}
The sign structure in the partition function is
\begin{equation}
(-1)^{P_\uparrow[c]+P_\downarrow[c]+S_{d\uparrow}[c]+S_{d\downarrow}[c]},
\end{equation}
where $P_\sigma[c]$ is the permutation parities of the $\sigma$-spinons due to the fermion statistics. It captures the fermion statistics in terms of the electrons because
\begin{equation}
P_\sigma[c]+S_{d\sigma}[c]\equiv P_\sigma^e[c]\pmod{2}
\end{equation}
where $P_\sigma^e[c]$ is the permutation parity of the spin-$\sigma$ \emph{electrons}.

On a bipartite lattice, its equivalence to Eq. (\ref{EqSSSB}) is demonstrated by
\begin{equation}
\tilde{S}_{\uparrow\downarrow}+S_{d\downarrow}+S_{h\uparrow}\equiv 0\pmod{2},
\end{equation}
where $\tilde{S}_{\uparrow\downarrow}=S_{\uparrow\downarrow}+C_{d\uparrow}-A_{d\uparrow}$ is the number of effective swaps between up- and down-spinons by treating doublons (holons) as a special type of up- (down-)spinons.

\end{widetext}
\bibliography{library,Books}
\end{document}